\documentclass[aps,prb,twocolumn,showpacs,preprintnumbers,superscriptaddress,amsmath,amssymb,longbibliography]{revtex4-1}

\usepackage{dcolumn}
\usepackage{bm}
\usepackage[normalem]{ulem}    
\usepackage{xcolor}
\usepackage{textcomp,gensymb}   
\usepackage{tikz}
\usepackage{physics}
\usepackage[pdftex,colorlinks=true,pdfstartview=FitV,linkcolor=blue,citecolor=blue,urlcolor=blue]{hyperref}
\usepackage{tabu}
\usepackage{soul}

\newcommand{\asb}{$\alpha$-Sb}
\newcommand{\sboro}{$\alpha$-Sb/Au(111)}
\newcommand{\aaa}{Au(111)}
\newcommand{\bfA}{\mathbf{A}}
\newcommand{\bfa}{\mathbf{a}}
\newcommand{\bfB}{\mathbf{B}}
\newcommand{\bfb}{\mathbf{b}}

\begin{document}

\title{Three-dimensional deformations in single-layer $\alpha$ antimonene and interaction with a Au(111) surface from first principles}
\author{Jos\'{e} de Jes\'{u}s Villalobos Castro}
\affiliation{Universit\'{e} Paris-Saclay, ONERA, CNRS, Laboratoire d'Etude des Microstructures (LEM), F-92322 Ch\^{a}tillon, France}

\author{Thomas Pierron}
\affiliation{Laboratoire de Physique et d'Etude des Mat\'{e}riaux (LPEM), ESPCI Paris-PSL Universtity, CNRS UMR8213, Sorbonne Universit\'{e}, 75005 Paris, France}

\author{Stephane Pons}
\affiliation{Laboratoire de Physique et d'Etude des Mat\'{e}riaux (LPEM), ESPCI Paris-PSL Universtity, CNRS UMR8213, Sorbonne Universit\'{e}, 75005 Paris, France}

\author{Johann Coraux}
\affiliation{Institut N\'{E}EL, CNRS and Universit\'{e} Joseph Fourier, BP166, F-38042 Grenoble Cedex 9, France}

\author{Lorenzo Sponza}
\email{lorenzo.sponza@onera.fr}
\affiliation{Universit\'{e} Paris-Saclay, ONERA, CNRS, Laboratoire d'Etude des Microstructures (LEM), F-92322 Ch\^{a}tillon, France}
\affiliation{European Theoretical Spectroscopy Facility (ETSF), 17 Sart-Tilman B-4000 Li\`{e}ge, Belgium}

\author{Sergio Vlaic}
\affiliation{Laboratoire de Physique et d'Etude des Mat\'{e}riaux (LPEM), ESPCI Paris-PSL Universtity, CNRS UMR8213, Sorbonne Universit\'{e}, 75005 Paris, France}


\begin{abstract}
Using density functional theory, we investigate the electronic structure of the $\alpha$ phase of an antimony monolayer (\asb{}) in its isolated form and in contact to the (111) surface of gold. We demonstrate that the isolated \asb{ }single-layer actually displays a slightly modulated puckering that stabilizes the monolayer, not a uniform one as often assumed. Moreover, it has dramatic consequences on the electronic band structure: the material is a semiconductor with low-dispersing bands near the Brillouin zone center. By further application of about 12\% strain on the armchair direction, a double-cone features develops wherein an electronic bandgap of $\sim$21~meV is found. When in contact with a Au(111) surface, a strong interaction with gold arises, as it appears clearly from (i) substantial atomic displacements compared to the isolated form, and (ii) hybridization of Sb and Au orbitals. The latter profoundly modifies the electronic band structure by strengthening the spin-orbit splitting of hybridized bands and spoiling the double-cone feature whose manipulation through substrate-induced strain appears therefore questionable, at least in the simulated epitaxial implementation.
\end{abstract}


\maketitle

A variety of elemental two-dimensional (2D) materials beyond graphene are known to date.
Materials belonging to the group Va
come with relatively low-symmetric non-flat atomic structures and host rich electronic properties often comprising a sizeable spin-orbit interaction (cfr. aresenene, atimonene and bismuthene).
Antimonene, in particular, is a synthetic material in some of its polymorphs\cite{Gibaja2016,Lei2016,Wu2017,Fortin2017} with strain-tunable electronic properties\cite{Akturk2015,Zhang2015} and a non-trivial band structure topology. \cite{Chuang2013,Huang2014,Zhao2015,Lu2016a,Lu2021,Lu2022}
The so-called $\alpha$ phase (\asb{}), a puckered structure similar to black phosphorus, is predicted to host anisotropic Dirac cones that are remarkably not pinned at high-symmetry points of the Brillouin zone.\cite{Lu2022,Lu2016a}

Heteroepitaxy, whereby a crystalline overlayer grows at the surface of a crystalline substrate, inherently implements interfacial stress.
Thus, it was argued that the position of the above-mentioned unpinned Dirac cones could be controlled through heteroepitaxial strain imposed by the substrate\cite{Zhao2015,Wang2015,Lu2016b,Mao2018}.
But such prediction was based essentially on simulations of isolated strained monolayers, whereas the influence of the substrate on the electronic properties of the overlayer can go beyond the simple strain because of electronic interactions occurring across the interface. 
Indeed, the extensive research on epitaxial graphene~\cite{Voloshina2012,Batzill2012} and other 2D materials\cite{Miwa2015,He2019,Mahatha2014} shows that, depending on the substrate, the pristine electronic properties can be preserved,\cite{Sutter2009,Nie2012} perturbed\cite{Pletikosi2009,Rusponi2012,Vincent2020} or even spoiled.\cite{Sutter2009b,Usachov2015,Wang2021}
The mechanism being very general, the effectiveness of heteroepitaxy as a strain engineering method for \asb{} deserves deeper investigation.

Growing \asb{} on the (001) surface of SnS\cite{Lu2022} and SnSe\cite{Shi2020,Shi2020b,Lu2022} seemed to confirm the prediction of movable Dirac-cones through epitaxial strain.
However, these substrates share the same crystal symmetry as \asb{}.
The Sb monolayer has been also grown on surfaces with different symmetries enabling other strain configurations. 
They include WTe$_2$,\cite{Shi2019} Bi$_2$Se$_3$,\cite{Hogan2019} MoTe$_2$,\cite{Shi2020b} the (111) surface of Ge\cite{Fortin2017} and the (111) surfaces of coinage metals.
In the latter cases, the Sb-metal affinity makes the synthesis quite complex because of the rich phase diagram which leads to  genuine 2D phases or surface alloys depending on the Sb dose and temperature.\cite{Niu2019,Zhou2019,Cantero2021,Zhang2022a}
For example, on Ag\cite{Zhang2022a} and Cu\cite{Zhang2022a} antimonene is reported to grow on top of a surface alloy which forms during synthesis, whereas on Au(111) the surface alloy disappears leaving place for an abrupt Au/Sb interface.\cite{Zhou2019,Cantero2021}
At the time being, no observation of Dirac cones has been reported on these substrates.

Here, we present a detailed density functional theory (DFT) analysis of the atomic structure and electronic band structure of single-layer \asb{}, subjected to strain and interacting with one of the coinage substrates, Au(111).
In the free-standing phase, we highlight the role of out-of-plane modulations that stabilise the free-standing layer and have remarkable effects on the band structure.
In the epitaxial configuration \sboro{}, we focus on distinguishing effects induced by the strain from those arising from the hybridization across the interface.
In this way, we answer the question whether heteroepitaxy on Au(111) is an effective method to strain-engineer the electronic properties of \asb{} and we point out other emergent phenomena stemming from the Sb/Au interaction.

\section{Computational and structural details}

We considered two distinct systems: the free-standing single-layer \asb{} and \sboro{}, i.e. epitaxial \asb{} on a \aaa{} surface.

\textit{Ab-initio} simulations have been carried out in the framework of DFT using the Quantum ESPRESSO package.~\cite{Giannozzi2009,Giannozzi2017}
The PBEsol~\cite{Perdew2008} exchange correlation potential with Grimme-D2~\cite{Grimme2006} van der Waals interactions has been chosen for an appropriate description of \sboro{}.
For consistency, the same potentials have been used also for the free-standing \asb{} monolayer.
Ultrasoft pseudopotentials with core corrections have been employed and the basis-set cutoff energies are 45~Ry for  \sboro{}
and 35~Ry for the free-standing one.
The Brillouin zone has been sampled with a $8\times5\times1$ Monkhorst-Pack grid in \sboro{} and the \aaa{} surface. 
A $11\times11\times1$ Monkhorst-Pack grid was used for the free-standing \asb{}.
In all cases, we have used simulation supercells comprising about 20 \AA{} of empty space in the vertical direction to avoid artefact interactions between replicas of the system.
Spin-orbit interaction (SOI) has been neglected in structural relaxation runs but included in all other calculation runs.

\begin{figure}
    \centering
    \includegraphics[width=0.99\linewidth]{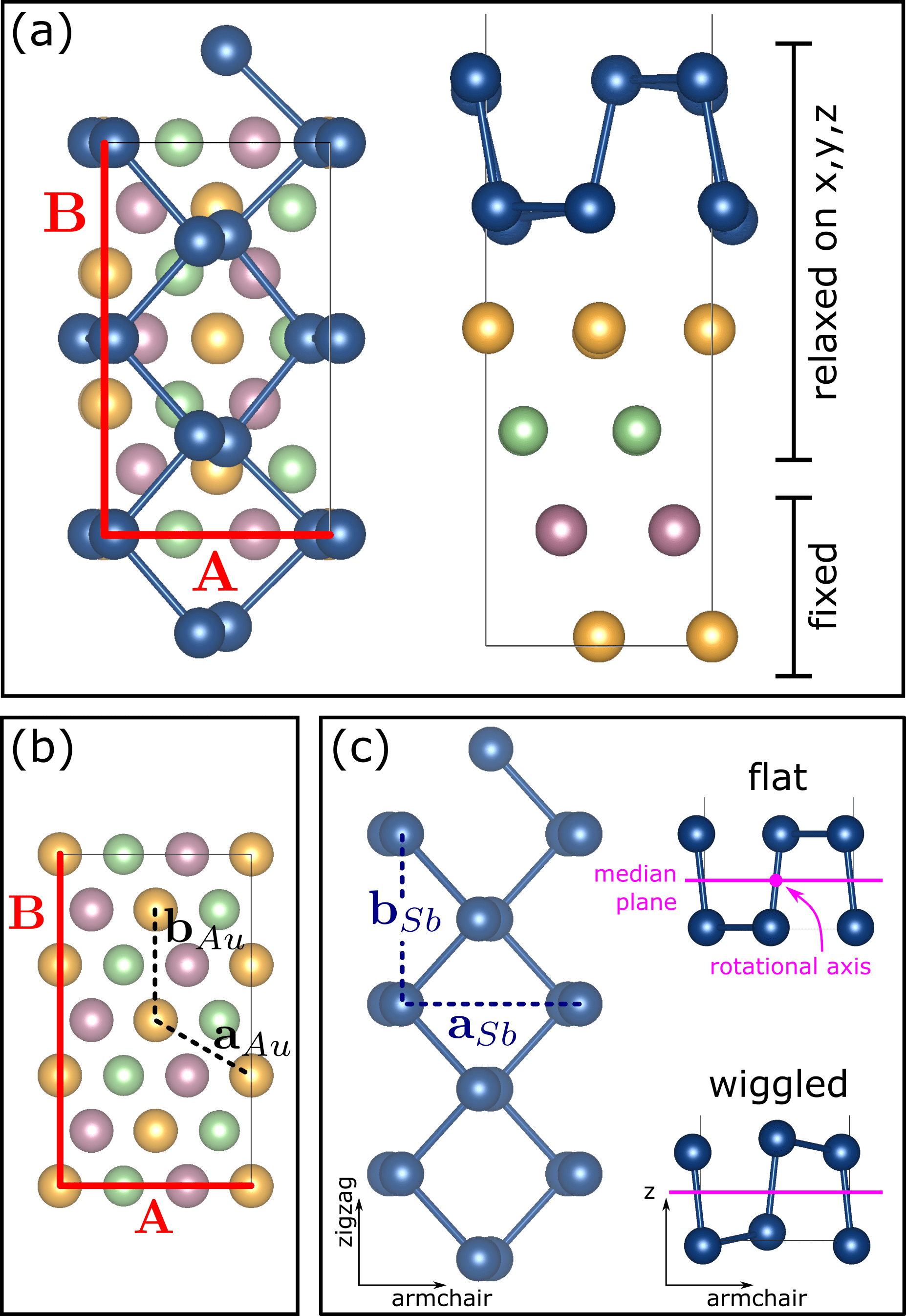}
    \caption{(a) Structure of \sboro{ }from above (left) and from the side (right). Blue spheres indicate Sb atoms; Yellow, green and purple spheres are Au atoms belonging to different layers. Red lines highlight the orthorhombic supercell vectors $\mathbf{A}$ and $\mathbf{B}$.
    Indications are given about the relaxation procedure. (b) Structure of the clean \aaa{ }surface. The orthorombic supercell vectors $\mathbf{A}$ and $\mathbf{B}$ are marked in red while the surface cell vectors $\bfa_\mathrm{Au}$ and $\bfb_\mathrm{Au}$ are in dashed black. (c) Left: Structure of the free-standing \asb{} with the unitary cell vectors $\mathbf{a}_\mathrm{Sb}$ and $\mathbf{b}_\mathrm{Sb}$ highlighted with dashed blue lines. Right: Sketch of the \emph{flat} and \emph{wiggled} structures with the horizontal rotational axis highlighted in magenta.}
    \label{fig:structures}
\end{figure}

Structural optimizations have been performed following a Broyden-Fletcher-Goldfarb-Shanno algorithm stopped when all force components on all atoms were lower than 10$^{-4}$ Ry/Bohr and total energy differences lower than 10$^{-6}$~Ry.
For the isolated \asb{} monolayer, full relaxation (cell parameters and all atomic positions) has been performed.
The final structure, reported in Figure~\ref{fig:structures}.c) has unitary cell parameters $a_\mathrm{Sb}=4.47$~\AA{ }and $b_\mathrm{Sb} =4.26$~\AA.
To model \sboro{} system, we considered that Sb and Au atoms do not alloy.
Hence, the Sb atoms are exclusively in the 2D overlayer and the Au substrate remains pure, which has been modelled with four layers of gold atoms. 
\aaa{} and \asb{} have dissimilar structure, yet a short-range coincidence is met experimentally\cite{Cantero2021,Zhang2022a} for an orthorhombic supercell having the in-plane cell parameters $A=5.01$~\AA{ }and $B=8.67$~\AA{ }. 
During relaxations, we kept these parameters fixed while
we allowed the Sb atoms and the top two layers of Au atoms to relax in the three Cartesian directions.
Tests carried out on a \aaa{} surface made of eight layers (not shown) ensured us that this procedure was safe.
The unitary cell vectors of the \aaa{} surface are $\bfa_\mathrm{Au} = s \begin{pmatrix} \sqrt{3}/2 \\ -1/2 \end{pmatrix}$ and $\bfb_\mathrm{Au} =  s \begin{pmatrix} 0 \\ 1 \end{pmatrix}$ with $s=2.89$~\AA.
They are related to the supercell parameters by the transformation matrix $\mathbf{M}$ 
\begin{equation}
    \begin{pmatrix} \bfA \\ \bfB \end{pmatrix} = \mathbf{M} \begin{pmatrix} \bfa_\mathrm{Au} \\ \bfb_\mathrm{Au} \end{pmatrix}
    \quad \text{with} \quad \mathbf{M} = \begin{pmatrix} 2 & 1 \\ 0 & 3 \end{pmatrix}
\end{equation}
as indicated by the experimental work of Cantero and coworkers.\cite{Cantero2021}

\section{Free-standing monolayer}


\begin{figure*}
    \centering
    \includegraphics[width=0.99\linewidth]{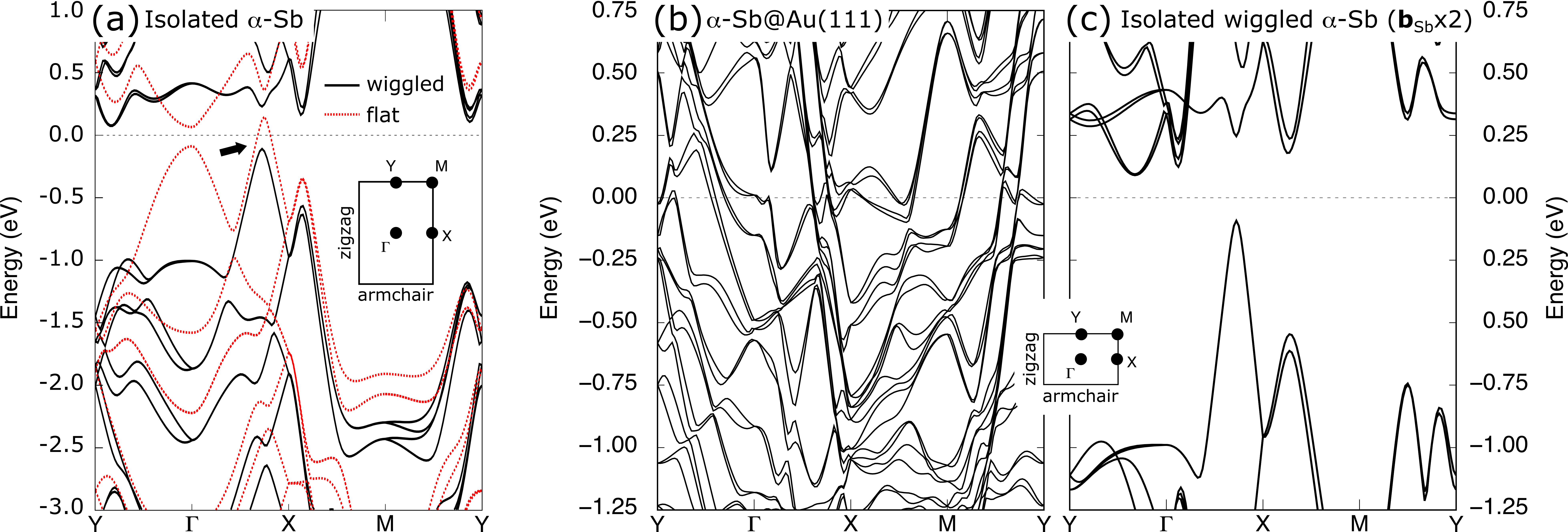}
    \caption{(a) Band structure of free-standing \asb{ }in the flat (red dashed) and the wiggled (solid black) configurations. The black arrow highlights the change from metallic to semiconducting behaviour. Inset: First Brillouin zone.
    (b) Band structure of \sboro{}. (c) Band structure of the free-standing wiggled \asb{} inside a supercell with a double $\bfb_\mathrm{Sb}$. Inset: First Brillouin zone of latter two systems.}
    \label{fig:sb-sboro}
\end{figure*}

The free-standing monolayer has a puckered structure where pairs of atoms alternate above and below a median plane (cfr. ball-and-stick models of Figure~\ref{fig:structures}c.
In some \emph{ab-initio} calculations, including recent ones,\cite{Lu2016a,Lu2022} the atoms forming a pair are assumed to lie on the same plane.
Thus, all atoms lie either on a top or a bottom plane which leads to the \emph{flat} structure depicted in the top-right corner of Figure~\ref{fig:structures}c.
Such a structure has a 180$^\circ$ rotational symmetry along an axis parallel to the zigzag direction passing through the middle of the bond between atoms of the two planes.
However, this symmetry will presumably break once the \asb{} monolayer interacts with a surface,\cite{Zhang2022a,Wang2015b,Zhao2015} for example when deposited on a substrate.
The simplest possible atomic configuration featuring such symmetry reduction is sketched in the bottom-right cartoon of Figure~\ref{fig:structures}c --- we dub it the \emph{wiggled} structure.
As a matter of fact, even in the free-standing monolayer, the wiggled structure is slightly more stable than the higher symmetry one.
We calculated the energy gain of wiggling to be of 0.02~eV/atom.
It turns out that this structure was already the one considered in early DFT calculations.\cite{Wang2015b}

What was not evident from these previous results was the extreme sensitivity of the electronic properties on the (faint) wiggling, which we will discuss below.
This symmetry breaking has several notable consequences.
On the structural side, neglecting the out-of-plane expansion associated to the wiggling imposes an artificial strain. 
By allowing for relaxation, the structure accommodates a shorter in-plane lattice constant $a_\mathrm{Sb}$, which passes from 4.65 to 4.47~\AA.
The electronic band structures of the flat and the wiggled configurations are shown in Figure~\ref{fig:sb-sboro}a.
The wiggling obviously lifts some of the bands' degeneracy, which is particularly visible along the $Y-\Gamma$ and the $M-Y$ directions.
A wiggling-induced flattening of the last occupied and first empty bands is also observed along the $Y-\Gamma-X$ path. 
The $\Gamma$ valleys are indeed very sensitive to strain and their dispersion almost vanishes when the structure is allowed to relax to its wiggled configuration.  
Another notable effect, though, is the widening of a gap between the conic features in the $\Gamma-X$ path, highlighted in the Figure by an arrow.
This is particularly remarkable as it makes the material change from metallic (flat) to semiconducting (wiggled).
This implies that controlling the wiggling in the free-standing (or weakly interacting) monolayer, \emph{e.g.} through uniaxial strain, could produce a metallic-to-semiconductor phase transition.

\section{supported monolayer}


In \sboro{} orthorhombic supercell (Figure~\ref{fig:structures}a),
the cell parameters\cite{Cantero2021} are $\bfA$=5.01~\AA{ }and $\bfB$=8.67~\AA. 
With this commensurability, the \asb{} monolayer is stretched anisotropically along two perpendicular directions by about 12\% ($\bfa_\mathrm{Sb}$, armchair direction) and by less than 2\% ($b_\mathrm{Sb}$, zigzag direction).
One should notice that \asb{} shares with other puckered monolayers a strong mechanical anisotropy that allows the armchair axis to sustain high levels of strain at a relatively low energy cost.\cite{Bienvenu2020}

The relaxed structure is charactertized by a vertical distance between the Sb layer and the Au surface varying between 2.40~\AA{} and 2.82~\AA{}, with an average Au-Sb distance of about 2.63~\AA{}.
This is quite small compared to the interlayer distance in graphite (about 3.35~\AA) and indicates that interactions occurring at the Sb/Au interface are stronger than van der Waals ones. 

The band structure of \sboro{} is reported in Figure~\ref{fig:sb-sboro}b. 
A comparison with the bands of the free-standing \asb{} discussed before is possible on the condition of drawing the bands of the isolated wiggled \asb{} within a supercell with double $\bfb_\mathrm{Sb}$, here reported in Figure~\ref{fig:sb-sboro}c.
At first sight, it may even seem hopeless to recognize any feature from the latter in the former.
First, gold's electrons obviously spawn a number of additional bands, which are all split by the SOI away from the $\Gamma$ point, like those of free-standing \asb{}, but with a different magnitude.
Second, the free-standing \asb{} monolayer is unstrained, unlike the deposited one.
In the following two subsections, we will develop a dedicated analysis to disentangle effects due to the strain on the Sb layer from effects due to the Sb-Au interaction.


\begin{figure*}
    \centering
    \includegraphics[width=0.98\linewidth]{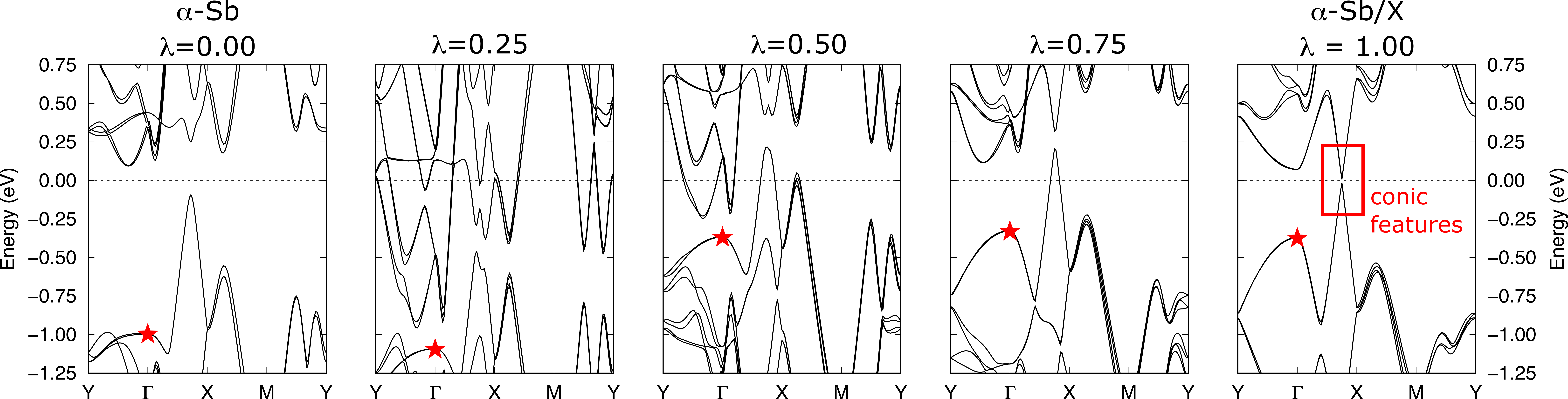}
    \caption{(a) Evolution of \asb{} band structure as a function of the linear interpolation parameter $\lambda$.
    As a guide for the eye, a red star highlights a notable dome feature in the valence band and a red rectangle highlights a conic feature in the \asb{}/X configuration (cfr. Figure~\ref{fig:cones}).
    Bands are aligned at the Fermi energy or the middle of the gap.}
    \label{fig:evoulution}
\end{figure*}   

\begin{figure}[b]
     \centering
     \includegraphics[width=0.98\linewidth]{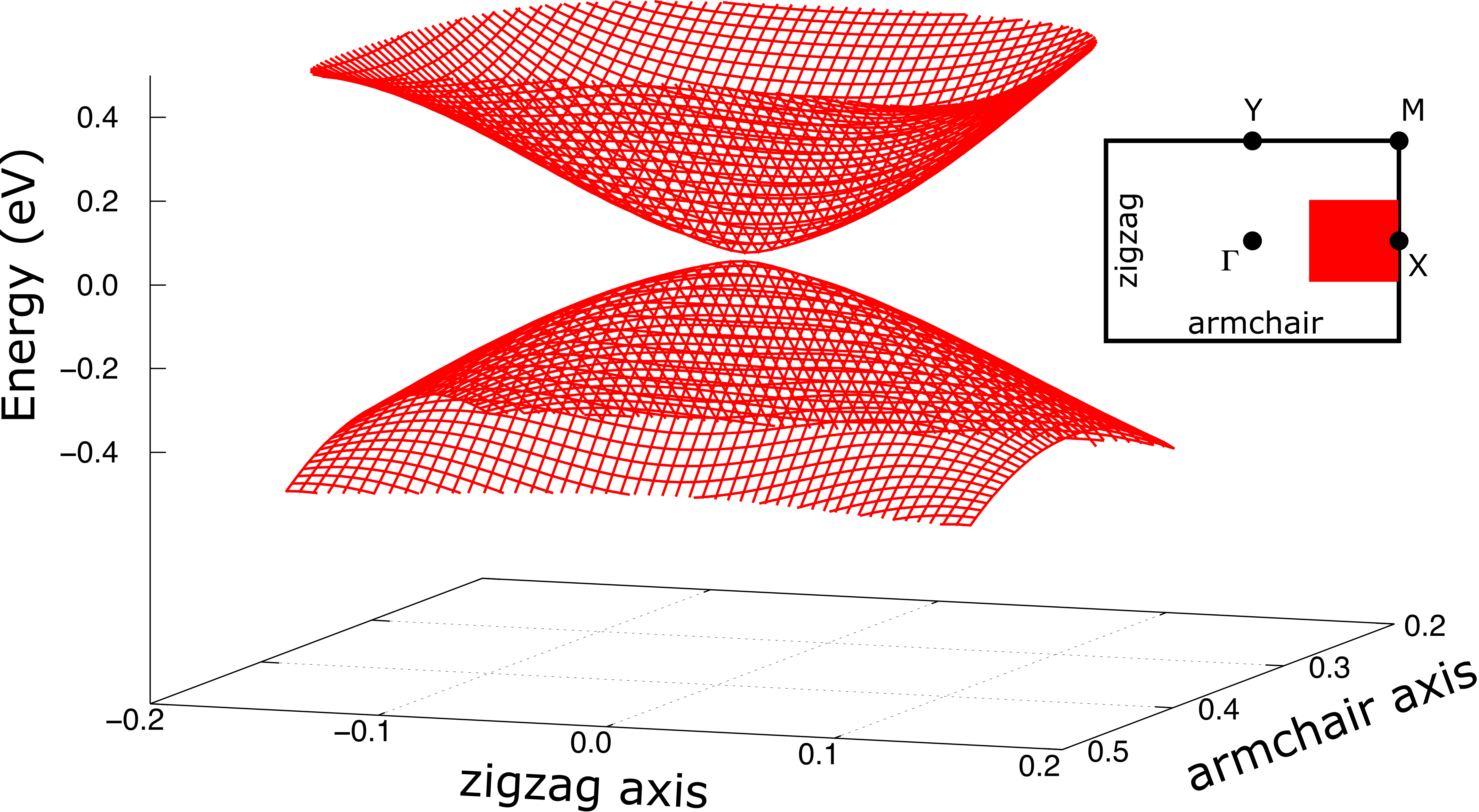}
     \caption{Three dimensional band dispersion of the conic feature highlighted in Figure~\ref{fig:evoulution}. The momentum coordinates $k_x$ (armchair) and $k_y$ (zigzag) are given in reciprocal lattice units. Inset: First Brillouin zone of the double-zigzag \asb{} monolayer with a red rectangle corresponding to the momentum domain of the 3D plot.}
    \label{fig:cones}
\end{figure}

\begin{figure*}
    \centering
    \includegraphics[width=0.99\linewidth]{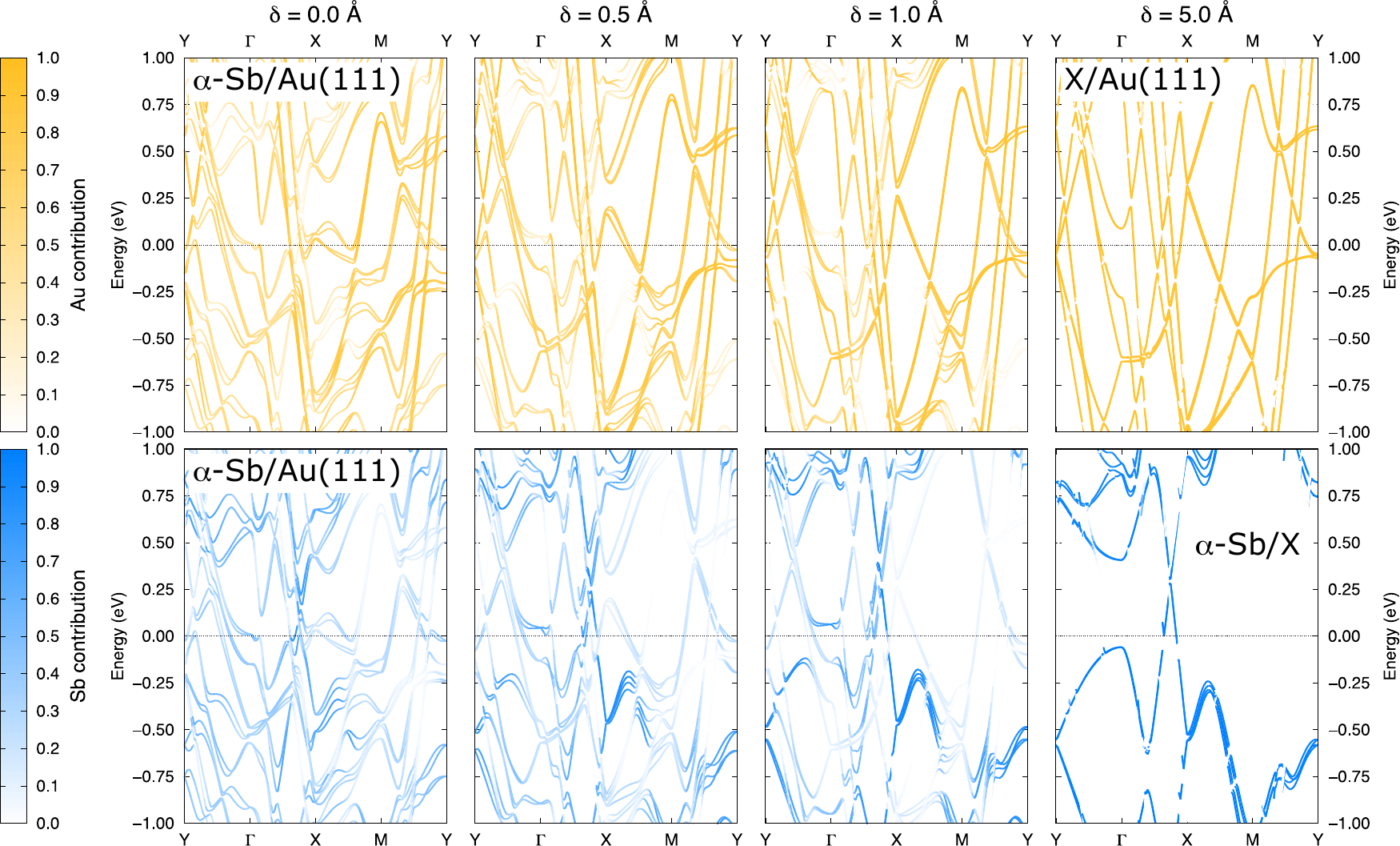}
    \caption{Electronic band structure as a function of the $\delta$ parameter (see text), calculated for \sboro{} and projected on the Au (top) and Sb (bottom) states.}
    \label{fig:fatbands_vs_distance}
\end{figure*}

\subsection{Strain effects}

We now remove all Au atoms (and their electrons) from the relaxed \sboro{} system, without performing any further relaxation of the structure.
We refer to this artificial configuration as ``\asb/X''.
Below, we scrutinize the effect of epitaxial strain by  progressively deforming the free-standing \asb{} into this configuration.
The deformation is parametrized with a single parameter, $\lambda$, used to linearly interpolate the atomic positions between the \asb{} configuration ($\lambda=0$) and the \asb{}/X one ($\lambda=1$).

Figure~\ref{fig:evoulution} displays the band structure calculated for five $\lambda$ values.
The valence bands in the $Y-\Gamma$ path are massively pushed to higher energies under substrate-induced strain, at $\lambda\geq0.25$. 
As a guide for the eye, a red star in the plot highlights two almost degenerate bands that are pushed to higher energies as $\lambda$ increases, and separated from the rest in $\Gamma$.
Concomitantly, a symmetric feature arises in the conduction band and eventually comes close to the Fermi energy (clearly identifiable at $\lambda=1$).
Remarkably, in the final step \asb/X, the monolayer develops  a characteristic double-cone feature in the $\Gamma-X$ path, with the two cones separated by a gap of only 21~meV, highlighted by a red frame in  Figure~\ref{fig:evoulution} and whose a 3D close-up is reported in Figure~\ref{fig:cones}.

Our analysis overall shows that epitaxial strain modifies sizeably the electronic structure of \asb.
Still, states of the \asb/X band structure are barely recognizable in \sboro{} band plot (Figure~\ref{fig:sb-sboro}b, which indicates that it epitaxial strain alone does not allow to predict the electronic properties of \sboro{}.
 We note that this conclusion is consistent with the relatively low Sb-Au distance highlighted above.

\subsection{Au-Sb hybridization}

Like we did with in the previous subsection, here we  also seek a single parameter that tunes the mixing of the Au and Sb states, \textit{i.e.} their hybridization.
This parameter is an artificial rigid shift $\delta$ of the Sb atoms in the $z$ direction, so that the $\delta=0$~\AA{} corresponds to the pristine \sboro{}.
Here again, no structural relaxation is performed when varying $\delta$, so the atomic arrangements of the Au surface and the Sb layer are the same as in the pristine \sboro{}, and hence the strain.

In Figure~\ref{fig:fatbands_vs_distance}, we report the band structure of configurations ranging from $\delta =0.0$~\AA{} (left panels) to $\delta=5.0$~\AA{} (right panels) with two intermediate steps in between that permit following the evolution.
Moreover, the intensity of each state is proportional to its projection on the Au (top panels) or Sb  (bottom panels) electronic components, the so-called ``fat band'' representation.

Let us start the discussion from the Au and Sb panels on the extreme right ($\delta=5$~\AA).
Note that $\delta$ is a rigid shift, so the actual Sb-Au distance in these plots ranges between 7.40~\AA{} and 7.82~\AA{}.
We observe that bands have neat Au or Sb character and no band appears in both plots, indicating that $\delta$=5~\AA{} is sufficiently high for the Au surface and the Sb layer to be isolated.
As a confirmation, in the Sb-projected plot (bottom-right) we retrieve the band structure of the \asb/X system (cfr. Figure~\ref{fig:sb-sboro}c.
On the other hand, the Au-projected states (top-right) correspond to the states of the Au surface deformed as if Sb were deposited on top. 
We label this system ``X/Au(111)''. 
It provides a guide for the eye that helps decipher the intricate network of the pristine \sboro{} bands at $\delta=0$~\AA.

Let us now consider the two $\delta=0$~\AA{} panels on the left.
Besides the intensity variations, in both plots we recognise the band structure of Figure~\ref{fig:sb-sboro}b.
The fact that all bands appear in both panels indicate clearly that basically all states have a mixed Au and Sb character, confirming the hybrid nature of the interface, at least inside the energy range we have chosen \emph{i.e.} close to Fermi.
As anticipated before, in both plots we recognize many features closely related to the X/Au(111) surface.
At variance, the Sb bands overall fade into a dense net of bands, with numerous crossings and small gaps opening and actually lose almost all resemblance with the isolated strained \asb/X band structure.

At intermediate $\delta$ values, we can track, for instance, the progressive fading of the conic features of $\alpha$-Sb and the hybridization spoiling its band structure especially in the $Y-\Gamma$ section.
We can also appreciate the progressive emergence of the interface states in both Au and Sb about the Fermi level.

An interesting observation is that the Sb-Au interaction has the effect of increasing the spin-orbit splitting of all bands.
This is actually surprising because Sb is lighter than Au.
A deeper analysis of this phenomenon will be presented in Ref.~\onlinecite{Pierron2024}.

\section{Conclusion}
We used DFT to study the electronic structure of a free-standing \asb{} monolayer and \sboro{}.

First of all, we highlight the importance of taking into account the vertical relaxation of Sb atoms in  the \asb{} monolayer.
The corresponding low-symmetry structure features a wiggling of the Sb pairs above and below the middle plane which, by accommodating planar strain, stabilizes further the monolayer.
Moreover, this symmetry breaking has dramatic consequences on the electronic structure of the isolated \asb{}, including the flattening of some bands in $\Gamma$ and, more importantly, the opening of a gap and a full filling of the valence band (semiconducting behaviour).

We also investigate the consequences of depositing \asb{} on a Au(111) surface, and thus consider an anisotropically strained commensurate structure that is experimentally relevant.
The Sb-Au equilibrium distance (ranging between 2.40~\AA{ }and 2.82~\AA) suggests that the interaction between Sb and Au is very strong.
Confirming this observation, signatures of the pristine Sb monolayer in the band structure of \sboro{} are barely recognizable, and this is mainly because of two reasons. 
The anisotropic strain, which is above 10\% in the armchair direction, translates in the appearance of a double-cone feature in the electronic band structure close to the Fermi level.
On top of this, the Au-Sb hybridization is strong enough to deeply modify the strained band structure. As a result, the double-cone feature is strongly altered and almost washed out by band crossings and small gap openings arising from a strong Au-Sb hybridization at the interface.

Finally, we also report that this Sb-Au mixing has the effect of globally enhancing the spin-orbit splitting of all bands --- a somehow counterintuitive finding, considering the small atomic weight of Sb atoms.

\begin{acknowledgements}
This research was founded, in part, by the French National Research Agency (ANR), grant numbers ANR-21-CE30-0043 (project SAGA) and ANR-21-CE09-0016 (project EXCIPLINT).
\end{acknowledgements}

%


\end{document}